\documentclass[aip,jcp,reprint,floatfix]{revtex4-1}
\usepackage{amsmath,amssymb}
\usepackage{epsfig}
\usepackage{graphicx}
\usepackage{dcolumn}
\usepackage{bbold}
\usepackage{natbib}
\usepackage{color}
\usepackage[colorlinks=true,citecolor={blue},urlcolor={blue}, linkcolor=blue]{hyperref}%
\begin{document}
\title{Search for parity and time reversal violating effects in HgH: Relativistic coupled-cluster study}%
\author{Sudip Sasmal}
\thanks{sudipsasmal.chem@gmail.com}
\affiliation{Electronic Structure Theory Group, Physical Chemistry Division, CSIR-National Chemical Laboratory, Pune, 411\,008, India}
\author{Himadri Pathak}%
\thanks{hmdrpthk@gmail.com}
\affiliation{Electronic Structure Theory Group, Physical Chemistry Division, CSIR-National Chemical Laboratory, Pune, 411\,008, India}
\author{Malaya K. Nayak}
\thanks{mk.nayak72@gmail.com}
\affiliation{Theoretical Chemistry Section, Bhabha Atomic Research Centre, Trombay, Mumbai 400\,085, India}
\author{Nayana Vaval}
\affiliation{Electronic Structure Theory Group, Physical Chemistry Division, CSIR-National Chemical Laboratory, Pune, 411\,008, India}
\author{Sourav Pal}
\affiliation{Department of Chemistry, Indian Institute of Technology Bombay, Powai, Mumbai 400\,076, India}
\begin{abstract}
The high effective electric field ($E_\mathrm{eff}$) experienced by the unpaired electron in an atom or a molecule is one
of the key ingredients in the success of electron electric dipole moment (eEDM) experiment and its precise calculation require a very accurate theory.
We, therefore, employed the Z-vector method in the relativistic coupled-cluster framework and
found that HgH has a very large $E_\mathrm{eff}$ value (123.2 GV/cm) which makes it a potential
candidate for the next generation eEDM experiment. Our study also reveals that it has a large scalar-pseudoscalar ${\mathcal{P,T}}$-violating
interaction constant, $W_\mathrm{s}$ = 284.2 kHz.
To judge the accuracy of the obtained results we have calculated parallel and perpendicular magnetic HFS constants and compared with the 
available experimental values. The results of our calculation are found to be in nice agreement with the experimental values.
Therefore, by looking at the HFS results we can say that both 
$E_\mathrm{eff}$ and $W_\mathrm{s}$ values are also very accurate.
Further, We have derived the relationship between these quantities and the ratio which will help to get model independent 
value of eEDM and S-PS interaction constant.
\end{abstract}
\maketitle
\section{Introduction \label{intro}}
In the quest of new physics, there have been an extensive search in order to observe the violation of parity (${\mathcal{P}}$) and time
reversal (${\mathcal{T}}$) symmetries \cite{ginges_2004}. 
An accurate measurement of electric dipole moment of an electron (eEDM) \cite{bernreuther_1991, commins_1999},
which arises due to the violation of both ${\mathcal{P}}$  and ${\mathcal{T}}$ is the most promising way to explore in this direction.
Although, the intensive search over the past half a century did not conclude in any final value of eEDM, however, it leads to achieve
a tremendous increase in the experimental sensitivity, and an upper bound limit of eEDM \cite{tho_edm}.
The enhancement of eEDM effects in heavy polar diatomic molecules is the main reason for the higher sensitivity
of modern eEDM experiment \cite{sushkov_1978, flambaum_1976}.
The sensitivity of the eEDM experiment using a heavy polar diatomic molecule depends on the molecule's permanent molecular
EDM \cite{mol_edm, hunter_1991}.
There are two main
sources of permanent molecular EDM of a paramagnetic molecule; the eEDM and the coupling interaction between the scalar-hadronic current
and the pseudoscalar electronic current.
However, in most of the calculation either eEDM or the scalar-pseudoscalar (S-PS) interaction is considered as the 
only possible source of permanent molecular EDM.
For example, the best upper bound limit of eEDM (d$_e$) and S-PS interaction constant (k$_s$) is obtained from the ThO experiment
by ACME collaboration \cite{tho_edm} where they have used the theoretically calculated value of effective electric field ($E_\mathrm{eff}$) and S-PS
P,T-odd interaction constant ($W_\mathrm{s}$) \cite{skripnikov_2013, skripnikov_2015}. In the calculation of d$_e$,
the S-PS coupling constant $k_s$, is assumed to be zero and vice versa, although both of these contribute
to the P,T-odd frequency shift in the experiment.
However, it is possible to get independent limit of d$_e$ and k$_s$ by using the results from two different experiments \cite{dzuba_2011}
and for this the accurate value of $E_\mathrm{eff}$, $W_\mathrm{s}$ and their ratio are very important.
Since, $E_\mathrm{eff}$ and $W_\mathrm{s}$ cannot be measured by any experimental technique,
therefore, these quantities have to be calculated by means of an accurate 
theoretical method, which can incorporate both the effects of relativity and electron correlation in an intertwined manner. \par
In this article, we focus on HgH molecule as it offers very high value of $E_\mathrm{eff}$ and $W_\mathrm{s}$ in its ground electronic state
($^{2}\Sigma$), which makes it a potential candidate for the future eEDM experiments.
The Z-vector method \cite{zvector_1989, sasmal_pra_rapid, sasmal_pbf} in the relativistic coupled-cluster formalism is used to calculate 
$E_\mathrm{eff}$ and $W_\mathrm{s}$ as it is the most reliable method for the calculation of ground state properties. 
However, high values of $E_\mathrm{eff}$ and $W_\mathrm{s}$ are not the only requirement for the precise measurement
of eEDM, but the molecule must be fully polarized with a low external electric field to fully utilize the $E_\mathrm{eff}$.
The large rotational constant and small dipole moment of HgH suggest that one needs to apply a much higher electric field
to polarize HgH in a spectroscopic experiment. However, Kozlov and Derevianko have suggested that it can be polarized easily 
in the matrix isolated solid state non-spectroscopic experiment \cite{kozlov_2006}, which also offers
2-3 orders of higher sensitivity than the current limit. Therefore, the detailed investigation of $E_\mathrm{eff}$ and $W_\mathrm{s}$ 
of HgH and their inter-relation is very important for the eEDM experiment based on HgH moecule. \par
The manuscript is organized as follows. Concise details of the calculated properties including a brief overview of the Z-vector method are
described in Sec. \ref{theory}. Computational details are given in Sec. \ref{comp}. We present our calculated
results and discuss about those in Sec. \ref{res_dis} before making final remark in Sec. \ref{conc}.
Atomic unit is used consistently unless stated. \par
\section{Theory \label{theory}}
\subsection{Properties \label{prop}}
The matrix element of $E_\mathrm{eff}$ is given by the following expression,
\begin{center}
\begin{eqnarray}
 E_\mathrm{eff} = |W_d \Omega|  = | \langle \Psi_{\Omega} | \sum_j^n \frac{H_d(j)}{d_e} | \Psi_{\Omega} \rangle |,
 \label{E_eff}
\end{eqnarray}
\end{center}
where $\Omega$ is the projection of total angular momentum along the molecular axis and $\Psi_{\Omega}$ is the
electronic wavefunction corresponding to $\Omega$ state. $n$ is the total number of electrons and 
H$_d$ is the interaction Hamiltonian of $d_e$ with internal electric field and is given by \cite{lindroth_1989, kozlov_1987, titov_2006}
\begin{eqnarray}
 H_d = 2icd_e \gamma^0 \gamma^5 {\bf \it p}^2 ,
\label{H_d}
\end{eqnarray}
where $\gamma$ are the usual Dirac matrices and {\bf \it p} is the momentum operator.
Another ${\mathcal{P,T}}$-odd interaction constant, $W_\mathrm{s}$, can be obtained by evaluating the following matrix element
\begin{eqnarray}
W_\mathrm{s}  = \frac{1}{\Omega k_s}\langle \Psi_{\Omega} | \sum_j^n H_{SP}(j) | \Psi_{\Omega} \rangle ,
\label{W_s}
\end{eqnarray}
where $k_s$ is the dimensionless electron-nucleus scalar-pseudoscalar coupling constant which is 
defined as Z$k_s$=(Z$k_{s,p}$+N$k_{s,n}$), where $k_{s,p}$ and $k_{s,n}$
are electron-proton and electron-neutron coupling constant, respectively.
The interaction Hamiltonian is defined as \cite{hunter_1991}
\begin{eqnarray}
H_{SP} = i\frac{G_F}{\sqrt{2}}Zk_s \gamma^0 \gamma^5 \rho_N(r) ,
\label{H_SP}
\end{eqnarray}
where $\gamma$ are the usual Dirac matrices, $\rho_N(r)$ is the nuclear charge density normalized to unity
and G$_F$ is the Fermi constant.
The calculation of the above matrix elements depends on the accurate wavefunction in the near nuclear region and the
standard way to determine the accuracy of the electronic wavefunction in that region is to compare the theoretically
calculated hyperfine structure (HFS) constant with the experimental value. The parallel ($A_{\|}$) and perpendicular
($A_{\perp}$) magnetic HFS constants of a diatomic molecule can be written as
\begin{eqnarray}
A_{\|(\perp)}= \frac{\vec{\mu_k}}{I\Omega} \cdot \langle \Psi_{\Omega} | \sum_i^n
\left( \frac{\vec{\alpha}_i \times \vec{r}_i}{r_i^3} \right)_{z(x/y)} | \Psi_{\Omega(-\Omega)}  \rangle,
\label{A_hfs}
\end{eqnarray}
where $\Omega$ represents the $z$-component of the total angular momentum of the diatomic molecule and 
it is 1/2 for the ground ($^{2}\Sigma$) state of HgH. \par
\subsection{Relation between independent limit of d$_e$ and k$_s$ with experimentally determined d$_e^{\mathrm{expt}}$}
Since, the dominant contribution of P,T-odd frequency shift comes from the eEDM and S-PS interaction, considering only those two
effects we can get the following relation \cite{tho_edm}
\begin{eqnarray}
 d_e E_\mathrm{eff} + \frac{W_\mathrm{s} k_s}{2} = \hbar \omega_{\!_{P,T}},
\end{eqnarray}
where $\hbar$ is the Planck's constant and $\omega_{\!_{P,T}}$ is the experimentally measured P,T-odd frequency shift.
\begin{eqnarray}
 \Longrightarrow d_e + \frac{W_\mathrm{s} k_s}{2 E_\mathrm{eff}} = \frac{\hbar \omega_{\!_{P,T}}}{E_\mathrm{eff}},
\end{eqnarray}
or,
\begin{eqnarray}
 d_e + \frac{k_s}{2R} = d_e^\mathrm{expt}|_{\!_{k_s=0}}.
 \label{rel}
\end{eqnarray}
Here, d$_e^{\mathrm{expt}}|_{\!_{k_s=0}}$ is the eEDM limit derived from the experimentally measured P,T-odd frequency shift
where k$_s$ is assumed to be 0 and R is defined as
\begin{eqnarray}
 R = \frac{E_\mathrm{eff}}{W_\mathrm{s}} .
\end{eqnarray}
Eq.\ref{rel} defines the interrelation of the independent limit of d$_e$, k$_s$ and experimentally determined d$_e^{\mathrm{expt}}$. \par
\subsection{Z-vector method \label{zvec}}
The relativistic coupled-cluster (CC) method is employed to calculate the molecular wavefunction. The CC energy is a
function of both the molecular orbital coefficients and determinantal coefficients in the expansion of a many electron
correlated wavefunction for a fixed nuclear geometry \cite{monkhorst_1977}. Thus, the CC energy derivative calculation needs to incorporate
the derivative of energy with respect to these two parameters in addition to the derivative of these two terms with
respect to the external field of perturbation.
However, the derivative of energy with respect to determinantal coefficients and the derivative of determinantal coefficients with respect
to external perturbation field can be included with the introduction of a perturbation independent linear equation whose solution yields
the Z-vector \cite{schafer_1984, zvector_1989}.

Computationally, Z-vector calculation in the CC framework is a three step process - (i) calculation of normal
coupled-cluster excitation operator (T), (ii) calculation of perturbation independent deexcitation operator $\Lambda$, and
(iii) the calculation of energy derivative using T and $\Lambda$. The form of the T and $\Lambda$ operators are given by
\begin{eqnarray}
 X=X_1+X_2+...+X_N=\sum_n^N X_n
\end{eqnarray}
with
\begin{eqnarray}
X_n= \sum\limits_{\stackrel{q_1<q_2\dots}{p_1<p_2\dots}}
x_{p_1p_2 \dots}^{q_1q_2\dots}{a_{q_1}^{\dagger}a_{q_2}^{\dagger} \dots a_{p_2} a_{p_1}} ,
\end{eqnarray}
where $x_{p_1p_2 \dots}^{q_1q_2\dots}$ are the cluster amplitudes corresponding 
to the cluster operator $X_m$. X is T when $p$($q$) are hole(particle) index and X is $\Lambda$ when $p$($q$) are particle(hole) index.
In the coupled-cluster single- and double- (CCSD) model, T is T$_1$ + T$_2$ and $\Lambda$ is $\Lambda_1$ + $\Lambda_2$.
The coupled-cluster T$_1$ and T$_2$ amplitudes can be calculated by solving the following equations,
\begin{eqnarray}
\langle \Phi_{i}^{a} | (H_Ne^T)_c | \Phi_0 \rangle = 0,\, \,\,
\langle \Phi_{ij}^{ab} | (H_Ne^T)_c | \Phi_0 \rangle = 0,
\label{cc_amplitudes}
\end{eqnarray}
where H$_N$ is the normal-ordered Dirac-Coulomb Hamiltonian and subscript $c$ means only the connected terms exist in the
contraction between H$_N$ and T. Size-extensivity is ensured by this connectedness.
The explicit equations for the amplitudes of $\Lambda_1$ and $\Lambda_2$ operators are give by
 \begin{eqnarray}
  \langle \Phi_0 |& [\Lambda (H_Ne^T)_c]_c | \Phi_{i}^{a} \rangle + \langle \Phi_0 | (H_Ne^T)_c | \Phi_{i}^{a} \rangle = 0,
  \label{lambda_1}
\end{eqnarray}
\begin{eqnarray}
 \langle \Phi_0 |& [\Lambda (H_Ne^T)_c]_c | \Phi_{ij}^{ab} \rangle + \langle \Phi_0 | (H_Ne^T)_c | \Phi_{i}^{a} \rangle \nonumber\\
 &  \times \langle \Phi_{i}^{a} | \Lambda | \Phi_{ij}^{ab} \rangle + \langle \Phi_0 | (H_Ne^T)_c | \Phi_{ij}^{ab} \rangle = 0.
\label{lambda_2}
\end{eqnarray}
Finally, the equation for energy derivative can be written as
\begin{eqnarray}
\Delta E' = \langle \Phi_0 | (O_Ne^T)_c | \Phi_0 \rangle + \langle \Phi_0 | [\Lambda (O_Ne^T)_c]_c | \Phi_0 \rangle,
\end{eqnarray}
where $O_N$ is the normal-ordered property operator. \par
\begin{table*}
\centering
\caption{Dipole Moment ($\mu$) (in Debye) and Magnetic HFS constants of HgH (in MHz)}
\begin{ruledtabular}
\newcommand{\mc}[3]{\multicolumn{#1}{#2}{#3}}
\begin{center}
\small
\begin{tabular}{lccccc}
× & × & \mc{2}{c}{$^{199}$Hg} & \mc{2}{c}{$^{201}$Hg}\\
\cline{3-4} \cline{5-6}
Basis &\, $\mu$ &\, A$_{\|}$ &\, A$_{\perp}$ &\, A$_{\|}$ &\, A$_{\perp}$ \\
\hline
TZ &\, 0.25 &\, 8371 &\, 6483 &\, $-$3090 &\, $-$2392 \\
QZ &\, 0.27 &\, 8440 &\, 6575 &\, $-$3116 &\, $-$2427 \\
Expt. &\, 0.47 &\, 7780(5)$^{a}$,\, 8200(60)$^{b}$ &\, 6200(3)$^{a}$,\, 6500(50)$^{b}$ &\, $-$2875(15)$^{a}$,\, $-$2980(40)$^{b}$ &\, $-$2275(10)$^{a}$,\, $-$2380(30)$^{b}$ \\
\end{tabular}
\end{center}
\label{hfs}
\end{ruledtabular}
$^{a}$measured in Ne matrices,
$^{b}$measured in Ar matrices.
\end{table*}
\section{Computational details \label{comp}}
The locally modified version of DIRAC10 program package \cite{dirac10} is used to solve the Dirac-Hartree-Fock equation where
the Dirac-Coulomb Hamiltonian is used.
On the other hand, the Z-vector method in the CCSD framework is used for the correlation treatment.
The wavefunction is four-component in nature and the small-component functions are linked to the large-components by the restricted
kinetic balance (RKB) condition \cite{dyall_2007}. Finite nucleus size is considered and the nuclear potential is calculated considering Gaussian charge
distribution \cite{visscher_1997}.
We have done two calculations - one with triple zeta (TZ) basis (dyall.cv3z for Hg \cite{dyall_5d} and cc-pCVTZ for H \cite{ccpcvxz_h})
and the other with quadruple zeta (QZ) basis (dyall.cv4z for Hg \cite{dyall_5d} and cc-pCVQZ for H \cite{ccpcvxz_h}). As the higher energy
virtual orbitals contribute very less
in the correlation calculations, the virtual orbitals whose energy exceeds 500 a.u. are removed from our calculations.
None of the occupied orbitals are frozen in our correlation calculation as the core
polarization effect plays a vital role for the type of properties of interest \cite{sasmal_pbf}. The experimental bond length of HgH
(1.766 \AA{}) \cite{herzberg_4} is used to calculate the properties in its ground state ($^{2}\Sigma_{1/2}$). \par
\begin{table}
\caption{$E_\mathrm{eff}$ (in GV/cm), $W_\mathrm{s}$ (in kHz) and the ratio of them (R = $E_\mathrm{eff}$/$W_\mathrm{s}$ in units of 10$^{18}$/e cm) of HgH.}
\begin{ruledtabular}
\newcommand{\mc}[3]{\multicolumn{#1}{#2}{#3}}
\begin{center}
\begin{tabular}{lcccccc}
Basis & \mc{2}{c}{$E_\mathrm{eff}$}&  \mc{2}{c}{$W_\mathrm{s}$}  & \mc{2}{c}{R}\\
\cline{2-3}  \cline{4-5}  \cline{6-7}
× & SCF & Z-vector & SCF & Z-vector & SCF & Z-vector\\
\hline
TZ & 106.8 & 123.3 & 241.2 & 284.3 & 107.1 & 104.9 \\
QZ & 106.9 & 123.2 & 241.7 & 284.2 & 106.9 & 104.8 \\
\end{tabular}
\end{center}
\label{pt_odd}
\end{ruledtabular}
\end{table}
\section{Results and discussion \label{res_dis}}
The accuracy of the property values given in Eq.\ref{E_eff} and Eq.\ref{W_s} can be determined by comparing the
theoretically obtained magnetic HFS constants
with the experimental values since all these matrix elements require an accurate wavefunction in the near nuclear region
of the heavy nucleus.
The parallel (A$_{\|}$) and perpendicular (A$_{\perp}$) magnetic HFS constant values of $^{199}$Hg and $^{201}$Hg in HgH
are presented in Table \ref{hfs}. The experimental values are taken from Ref. \cite{hgh_hfs}, where
Stowe {\it et al} measured the magnetic HFS constant of HgH trapped in neon and argon matrices at 4K by electron spin resonance
study. The agreement of our calculated A$_{\|}$ and A$_{\perp}$ results with the experimental values shows that the wavefunction evaluated
in Z-vector method is very accurate in the near nuclear region and thus it also shows the reliability of our calculated
$E_{\mathrm{eff}}$ and $W_\mathrm{s}$ values.
Further, we have calculated molecular-frame dipole moment ($\mu$) of the HgH molecule using same Z-vector method.
The obtained ($\mu$) values are 0.25 D and 0.27 D in TZ and QZ basis, respectively. These results are also compiled in the same table 
and compared with the available experimental value \cite{nedelec_1989}.
However, the experimental $\mu$ value of HgH reported in Ref. \cite{nedelec_1989} was measured with an unusual and indirect way
and the value is also given without any experimental uncertainty.
\par
In Table \ref{pt_odd}, we present the $E_\mathrm{eff}$ and $W_\mathrm{s}$ values of our calculation.
The $E_\mathrm{eff}$
value of HgH obtained in QZ basis is 123.2 GV/cm. This result shows that HgH is one of those diatomic molecules which have the
largest effective electric field.
Previously, Kozlov calculated the $E_\mathrm{eff}$ of HgH by using a semiempirical method and the value found to be 79 GV/cm. On the other
hand, we have used an {\it ab initio} (Z-vector method in the relativistic CCSD framework) method with sufficiently large basis sets
(TZ and QZ) to calculate the properties of HgH which makes our calculated values more reliable.
Our calculated $W_\mathrm{s}$ value in QZ basis is 284.2 kHz. This large value of $W_\mathrm{s}$ suggests that the S-PS interaction can
contribute a significant amount to the permanent molecular EDM.
These characteristics of HgH make it an important player in the field of eEDM search.
The ratio (R) of $E_\mathrm{eff}$ to $W_\mathrm{s}$ is also calculated as this is a very important quantity to obtain
the model independent limit of $d_e$ and $k_s$ as suggested by Dzuba {\it et al} \cite{dzuba_2011}. They suggest that this ratio would be
constant for a particular heavy nucleus for the following reasons: (i) these types of property mainly depend on the core
(near nuclear region) electronic wavefunction and in this short distance the one electron Dirac equation becomes identical
for all single-electron states
with the given angular momenta;
(ii) the main contribution of
these types of properties comes from the $s_{\!_{1/2}}$-$p_{\!_{1/2}}$ matrix elements and thus the many-body effects like core polarization
has a very little effect on the ratio R. Our calculated values of R in QZ basis are 106.9 and 104.8 in units of 10$^{18}$/e cm in the SCF and
in the Z-vector calculations, respectively. These results support the previous argument as the correlation treatment changes the value of
R only by 2 units. These values are very close to the value obtained by Dzuba {\it et al} (112.5 in the same unit) \cite{dzuba_2011} where they
used an analytic expression to evaluate this ratio.
By putting the value of R in Eq.\ref{rel}, we can get the following relation
\begin{equation}
 d_e + 4.77 \times 10^{-21} k_s = d_e^\mathrm{expt}|_{\!_{k_s=0}} ,
 \label{relation}
\end{equation}
where $d_e^\mathrm{expt}|_{\!_{k_s=0}}$ is the eEDM limit derived from the P,T-odd energy shift of HgH experiment
at the limit of k$_s$ = 0. \par
\begin{table}[ht]
\caption{ Convergence pattern of A$_{\|}$ of $^{199}$Hg in HgH, $\mu$, $E_{\mathrm{eff}}$ and $W_{\mathrm{s}}$ of HgH as a function of virtual orbitals }
\begin{ruledtabular}
\newcommand{\mc}[3]{\multicolumn{#1}{#2}{#3}}
\begin{center}
\begin{tabular}{lccccr}
Cutoff & Virtual & $\mu$ & A$_{\|}$ & $E_{\mathrm{eff}}$ & $W_{\mathrm{s}}$ \\
(a.u.) & spinor & (D) & (MHz) & (GV/cm) & (kHz) \\
\hline
100 & 175 & 0.168 & 7911 & 115.6 & 260.5 \\
200 & 199 & 0.169 & 7918 & 115.7 & 260.6 \\
400 & 207 & 0.169 & 7977 & 116.8 & 263.0\\
500 & 231 & 0.170 & 7980 & 116.8 & 263.1 \\
1000 & 239 & 0.170 & 8015 & 117.5 & 264.6 \\
no cutoff & 355 & 0.170 & 8069 & 118.5 & 266.9 \\
\end{tabular}
\end{center}
\end{ruledtabular}
\label{virtual}
\end{table}
There are three main possible sources of error associated with our calculation - (i) basis set incompleteness, (ii) cutoff used for
virtual orbitals in the correlation calculations and (iii) higher order correlation effect. The error associated with the
incompleteness of the basis set can be estimated by comparing TZ and QZ basis calculations.
The difference between the results calculated in TZ and QZ basis for both $E_\mathrm{eff}$ and $W_\mathrm{s}$ is less than 0.1\%.
We have done a series of calculations (compiled in Table \ref{virtual})
to estimate the error associated with the restriction of correlation space by neglecting higher energy virtual orbitals.
In this calculation, the $E_{\mathrm{eff}}$ and $W_{\mathrm{s}}$ values are calculated by employing double zeta (DZ) basis (dyall.cv2z
for Hg \cite{dyall_5d} and cc-pCVDZ for H \cite{ccpcvxz_h}) in the Z-vector method with different cutoff of virtual orbitals.
The difference of calculated values using 500 a.u. cutoff for virtual orbitals and using all virtual orbitals in
correlation calculation are 1.7 GV/cm and 3.8 kHz for $E_{\mathrm{eff}}$ and $W_{\mathrm{s}}$, respectively.
Therefore, if we use 500 a.u. as cutoff for the virtual orbitals, then the
associated errors in both $E_\mathrm{eff}$ and $W_\mathrm{s}$ values are about 1.4\%.
The effect of higher order correlation terms can be estimated by comparing our CCSD results with 
CCSD with partial triples (CCSD(T)) or with full configuration interaction (FCI) calculations. These types of calculations are 
very expensive and beyond the scope of the present study. However, from our experience we can comment that the error
associated with this effect will be within 3.5\%. Although these three effects are intertwined in nature, assuming linearity,
we estimate our results are correct within 5\% uncertainty. It is worth to mention that our results are free from the error
associated with the effect of core polarization since all the electrons are correlated in our calculation. \par
From Table \ref{hfs}, \ref{pt_odd} and \ref{virtual}, we can see that the HFS constants, $E_\mathrm{eff}$ and $W_\mathrm{s}$ values
are following the same trend and thus, we can comment that the calculated $E_\mathrm{eff}$ and $W_\mathrm{s}$ values are very accurate.
The above results suggest that HgH can be a potential candidate for the future eEDM experiment.
However, there are other factors that need to be considered for an eEDM experiment.
The ground state of HgH is a $^{2}\Sigma$ state. It has no orbital angular momentum contribution to the magnetic moment and
thus it cannot cancels the unpaired electron's spin angular momentum contribution to the magnetic moment unlike PbF and ThO.
For this reason the $^{2}\Sigma$ state of HgH has a higher $g$-factor compared to $^{2}\Pi_{1/2}$ state of PbF \cite{gfactor_pbf}
or $^{3}\Delta_1$ state of ThO \cite{vutha_2010, petrov_2014}. Thus, it can give a significant magnetic noise in the spectroscopic eEDM experiment.
Being a $^{2}\Sigma$ state, there are no $\Omega$-doublets \cite{herzberg_1} available for the ground state of HgH which can be used
as a comagnetometer state \cite{comagnetometer, demille_2000, petrov_2014} that can suppress some systematic
errors though it is possible to find the sets of
``internal comagnetometer'' states with the proper combinations of different rotational states \cite{prasanna_2015}.
Even though HgH has a high $E_{\mathrm{eff}}$, one needs to fully polarize the molecule in an external laboratory electric
field for the maximal utilization of $E_{\mathrm{eff}}$. But the large rotational constant \cite{veseth_1972} and
small molecular dipole moment \cite{nedelec_1989} of HgH suggest
that enormous amount of external electric field is required to fully polarize the HgH molecule.
On the contrary, as suggested by Kozlov {\it et al} \cite{kozlov_2006}, HgH can be polarized easily in the 
matrix isolated solid state non-spectroscopic experiment. They also argued that in this method, it is possible to
improve the eEDM limit by 2-3 orders higher than the current limit.
Therefore, considering these facts, we can comment that HgH can be a potential candidate for eEDM experiment and the solid state 
non-spectroscopic experimental technique would be best suitable for it. \par
\section{Conclusion \label{conc}}
In summary, we have performed the precise calculation of $E_{\mathrm{eff}}$ and $W_{\mathrm{s}}$ of HgH molecule in its open-shell
ground state using the most reliable Z-vector method in the relativistic coupled-cluster framework.
The outcome of our study reveals that HgH has one of the highest $E_{\mathrm{eff}}$ and $W_{\mathrm{s}}$ known for
the polar diatomic molecule. On the other hand, HgH can be polarized easily using a solid state non-spectroscopic technique.
Thus, the combination makes HgH a very important candidate for the next generation eEDM experiment. \par
\section*{Acknowledgrments}
We thank Prof. Victor Flambaum, Prof. Mikhail Kozlov and Dr. Amar Vutha for their valuable comments and insights in this work.
Authors acknowledge a grant from CSIR 12th Five Year Plan project on Multi-Scale Simulations of Material (MSM)
and the resources of the Center of Excellence in Scientific Computing at CSIR-NCL. S.S. and H.P acknowledge the CSIR
for their fellowship.
S.P. acknowledges funding from J. C. Bose Fellowship grant of Department of Science and Technology (India).
\section*{References}

\begin{thebibliography}{10}

\bibitem{ginges_2004}
{\sc J.~Ginges} and {\sc V.~Flambaum},
\newblock {\em Physics Reports} {\bf 397}, 63  (2004).

\bibitem{bernreuther_1991}
{\sc W.~Bernreuther} and {\sc M.~Suzuki},
\newblock {\em Rev. Mod. Phys.} {\bf 63}, 313 (1991).

\bibitem{commins_1999}
{\sc E.~D. Commins},
\newblock {\em Advances In Atomic, Molecular, and Optical Physics} {\bf 40}, 1
  (1999).

\bibitem{tho_edm}
{\sc J.~Baron}, {\sc W.~C. Campbell}, {\sc D.~DeMille}, {\sc J.~M. Doyle}, {\sc
  G.~Gabrielse}, {\sc Y.~V. Gurevich}, {\sc P.~W. Hess}, {\sc N.~R. Hutzler},
  {\sc E.~Kirilov}, {\sc I.~Kozyryev}, {\sc B.~R. O’Leary}, {\sc C.~D.
  Panda}, {\sc M.~F. Parsons}, {\sc E.~S. Petrik}, {\sc B.~Spaun}, {\sc A.~C.
  Vutha}, and {\sc A.~D. West},
\newblock {\em Science} {\bf 343}, 269 (2014).

\bibitem{sushkov_1978}
{\sc O.~Sushkov} and {\sc V.~Flambaum},
\newblock {\em Journal of Experimental and Theoretical Physics} {\bf 48}, 608
  (1978).

\bibitem{flambaum_1976}
{\sc V.~V. Flambaum},
\newblock {\em Sov. J. Nucl. Phys.} {\bf 24}, 199 (1976).

\bibitem{mol_edm}
The EDM of a polar diatomics due to the different number of proton in two
  nucleus and uneven distribution of electron density is not really be called
  as permanent molecular EDM. This is because the Stark shift induced by this
  EDM in an external laboratory field does not increase linearly rather
  quadratically with the electric field in the weak field limit. Thus strictly
  speaking, a polar diatomic molecule cannot have a permanent molecular EDM
  unless there is a violation of both P and T. \cite{hunter_1991}.

\bibitem{hunter_1991}
{\sc L.~R. Hunter},
\newblock {\em Science} {\bf 252}, 73 (1991).

\bibitem{skripnikov_2013}
{\sc L.~Skripnikov}, {\sc A.~Petrov}, and {\sc A.~Titov},
\newblock {\em The Journal of chemical physics} {\bf 139}, 221103 (2013).

\bibitem{skripnikov_2015}
{\sc L.~Skripnikov} and {\sc A.~Titov},
\newblock {\em The Journal of chemical physics} {\bf 142}, 024301 (2015).

\bibitem{dzuba_2011}
{\sc V.~A. Dzuba}, {\sc V.~V. Flambaum}, and {\sc C.~Harabati},
\newblock {\em Phys. Rev. A} {\bf 84}, 052108 (2011).

\bibitem{zvector_1989}
{\sc E.~A. Salter}, {\sc G.~W. Trucks}, and {\sc R.~J. Bartlett},
\newblock {\em The Journal of Chemical Physics} {\bf 90}, 1752 (1989).

\bibitem{sasmal_pra_rapid}
{\sc S.~Sasmal}, {\sc H.~Pathak}, {\sc M.~K. Nayak}, {\sc N.~Vaval}, and {\sc
  S.~Pal},
\newblock {\em Phys. Rev. A} {\bf 91}, 030503 (2015).

\bibitem{sasmal_pbf}
{\sc S.~Sasmal}, {\sc H.~Pathak}, {\sc M.~K. Nayak}, {\sc N.~Vaval}, and {\sc
  S.~Pal},
\newblock {\em The Journal of Chemical Physics} {\bf 143}, 084119 (2015).

\bibitem{kozlov_2006}
{\sc M.~G. Kozlov} and {\sc A.~Derevianko},
\newblock {\em Phys. Rev. Lett.} {\bf 97}, 063001 (2006).

\bibitem{lindroth_1989}
{\sc E.~Lindroth}, {\sc B.~Lynn}, and {\sc P.~Sandars},
\newblock {\em Journal of Physics B: Atomic, Molecular and Optical Physics}
  {\bf 22}, 559 (1989).

\bibitem{kozlov_1987}
{\sc M.~G. Kozlov}, {\sc V.~Fomichev}, {\sc Y.~Y. Dmitriev}, {\sc L.~N.
  Labzovsky}, and {\sc A.~V. Titov},
\newblock {\em Journal of Physics B: Atomic and Molecular Physics} {\bf 20},
  4939 (1987).

\bibitem{titov_2006}
{\sc A.~V. Titov}, {\sc N.~S. Mosyagin}, {\sc A.~N. Petrov}, {\sc T.~A. Isaev},
  and {\sc D.~P. DeMille},
\newblock {\em Progr. Theor. Chem. Phys.} {\bf 15}, 253 (2006).

\bibitem{monkhorst_1977}
{\sc H.~J. Monkhorst},
\newblock {\em Int. J. Quantum Chem.} {\bf 12}, 421 (1977).

\bibitem{schafer_1984}
{\sc N.~C. Handy} and {\sc H.~F. Schaefer},
\newblock {\em The Journal of Chemical Physics} {\bf 81}, 5031 (1984).

\bibitem{dirac10}
{DIRAC}, a relativistic ab initio electronic structure program, Release
  {DIRAC10} (2010), written by T.~Saue, L.~Visscher and H.~J.~{\relax
  Aa}.~Jensen, with contributions from R.~Bast, K.~G.~Dyall, U.~Ekstr{\"o}m,
  E.~Eliav, T.~Enevoldsen, T.~Fleig, A.~S.~P.~Gomes, J.~Henriksson,
  M.~Ilia{\v{s}}, Ch.~R.~Jacob, S.~Knecht, H.~S.~Nataraj, P.~Norman, J.~Olsen,
  M.~Pernpointner, K.~Ruud, B.~Schimmelpfennig, J.~Sikkema, A.~Thorvaldsen,
  J.~Thyssen, S.~Villaume, and S.~Yamamoto (see
  \url{http://www.diracprogram.org}).

\bibitem{dyall_2007}
{\sc K.~Faegri~Jr} and {\sc K.~G. Dyall},
\newblock {\em Introduction to relativistic quantum chemistry},
\newblock Oxford University Press, USA, 2007.

\bibitem{visscher_1997}
{\sc L.~Visscher} and {\sc K.~Dyall},
\newblock {\em Atomic Data and Nuclear Data Tables} {\bf 67}, 207  (1997).

\bibitem{dyall_5d}
{\sc K.~G. Dyall},
\newblock {\em Theoretical Chemistry Accounts} {\bf 112}, 403 (2004).

\bibitem{ccpcvxz_h}
{\sc T.~H. Dunning},
\newblock {\em The Journal of Chemical Physics} {\bf 90}, 1007 (1989).

\bibitem{herzberg_4}
{\sc K.~Huber} and {\sc G.~Herzberg},
\newblock {\em Constants of Diatomic Molecules, Vol. 4, Molecular Spectra and
  Molecular Structure},
\newblock Van Nostrand Reinhold, New York, 1979.

\bibitem{hgh_hfs}
{\sc A.~C. Stowe} and {\sc L.~B. Knight~Jr},
\newblock {\em Molecular Physics} {\bf 100}, 353 (2002).

\bibitem{nedelec_1989}
{\sc O.~Nedelec}, {\sc B.~Majournat}, and {\sc J.~Dufayard},
\newblock {\em Chemical Physics} {\bf 134}, 137  (1989).

\bibitem{gfactor_pbf}
{\sc L.~V. Skripnikov}, {\sc A.~N. Petrov}, {\sc A.~V. Titov}, {\sc R.~J.
  Mawhorter}, {\sc A.~L. Baum}, {\sc T.~J. Sears}, and {\sc J.-U. Grabow},
\newblock {\em Phys. Rev. A} {\bf 92}, 032508 (2015).

\bibitem{vutha_2010}
{\sc A.~C. Vutha}, {\sc W.~C. Campbell}, {\sc Y.~V. Gurevich}, {\sc N.~R.
  Hutzler}, {\sc M.~Parsons}, {\sc D.~Patterson}, {\sc E.~Petrik}, {\sc
  B.~Spaun}, {\sc J.~M. Doyle}, {\sc G.~Gabrielse}, and {\sc D.~DeMille},
\newblock {\em Journal of Physics B: Atomic, Molecular and Optical Physics}
  {\bf 43}, 074007 (2010).

\bibitem{petrov_2014}
{\sc A.~Petrov}, {\sc L.~Skripnikov}, {\sc A.~Titov}, {\sc N.~Hutzler}, {\sc
  P.~Hess}, {\sc B.~O'Leary}, {\sc B.~Spaun}, {\sc D.~DeMille}, {\sc
  G.~Gabrielse}, and {\sc J.~Doyle},
\newblock {\em Physical Review A} {\bf 89}, 062505 (2014).

\bibitem{herzberg_1}
{\sc G.~Herzberg},
\newblock {\em Molecular Spectra and Molecular Structure, Vol. 1, Spectra of
  Diatomic Molecules},
\newblock Van Nostrand Reinhold, New York, 1950.

\bibitem{comagnetometer}
The comagnetometer state \cite{demille_2000, petrov_2014} means two states that
  have comparable magnetic moment but opposite polarizibility. In spectroscopic
  eEDM experiment, the existance of a comagnetometer state is very important
  because it allows one to reverse the $E_{eff}$ spectroscopically (i.e., by
  populating the molecules in different comagnetometer state), without
  reversing the external laboratory electric field, which is used to polarize
  the molecule.

\bibitem{demille_2000}
{\sc D.~DeMille}, {\sc F.~Bay}, {\sc S.~Bickman}, {\sc D.~Kawall}, {\sc
  D.~Krause~Jr}, {\sc S.~Maxwell}, and {\sc L.~Hunter},
\newblock {\em Physical Review A} {\bf 61}, 052507 (2000).

\bibitem{prasanna_2015}
{\sc V.~S. Prasannaa}, {\sc A.~C. Vutha}, {\sc M.~Abe}, and {\sc B.~P. Das},
\newblock {\em Phys. Rev. Lett.} {\bf 114}, 183001 (2015).

\bibitem{veseth_1972}
{\sc L.~Veseth},
\newblock {\em Journal of Molecular Spectroscopy} {\bf 44}, 251  (1972).

\end{thebibliography}


\end{document}